# Modeling the recovery phase of extreme geomagnetic storms

C. Cid[1], J. Palacios[1], E. Saiz[1], Y. Cerrato[1], J. Aguado[2], A. Guerrero[1]

[1]Space Research Group - Space Weather, Departamento de Física y Matemáticas, Universidad de Alcalá, Alcalá de Henares, Spain.

[2]E. U. Cardenal Cisneros. Universidad de Alcalá, Alcalá de Henares, Spain.

Corresponding author: Consuelo Cid, Space Research Group - Space Weather, Departamento de Física y Matemáticas, Universidad de Alcalá, 28871 Alcalá de Henares (Madrid), SPAIN (consuelo.cid@uah.es).

**Key points:**

- The hyperbolic function proposed by Aguado et al. [2011] is checked during the most severe storms ever registered, as the Carrington event.

- A Local Disturbance Index, the $LDi$, is introduced to provide information of the disturbance during geomagnetic storms at a local scale when $Dst$ is not available.

- The linear relationship between the time that takes the magnetosphere to recover and the intensity of the storm is revised

**Index Terms:** Space Weather Models; Magnetic Storms; Magnetospheric configuration and dynamics

**Abstract.** The recovery phase of the largest storms ever recorded has been studied. These events provide an extraordinary opportunity for two goals: (1) to validate the hyperbolic model by *Aguado et al.* [2010] for the recovery phase after disturbances as severe as the Carrington event, or that related to the Hydro-Quebec blackout in March 1989, and (2) to check whether the linear relationship between the recovery time and the intensity of the storm still complies. Our results reveal the high accuracy of the hyperbolic decay function to reproduce the recovery phase of the magnetosphere after an extreme storm. Moreover, the characteristic time that takes the magnetosphere to recover depends in an exponential way on the intensity of the storm, as indicated by the relationship between the two parameters involved in the hyperbolic decay. This exponential function can be approached by a linear function when the severity of the storm diminishes.

## 1. Introduction

The term 'intense storm' is commonly used for a storm when the *Dst* index reaches -100 nT [*Gonzalez et al.*, 1994]. However, when it exceeds -250 nT, it is labeled as 'extreme storm', 'severe storm', 'great magnetic storm' or 'superstorm' [*Tsurutani et al.*, 1992; *Gonzalez et al.*, 2002; Echer et al., 2008]. The analysis of extreme geomagnetic storms, as a natural hazard, is used to describe a worst reasonable case scenario and the potential vulnerabilities and consequences. Many such efforts are operational measures relying on adequate warning.

The most severe geomagnetic storm of the past thirty years, the 1989 storm responsible for the Hydro-Quebec power blackout, registered a *Dst* minimum value of -640 nT. Although no recorded geomagnetic storm since 1932 exceeded -760 nT [*Cliver and*

*Svalgaard*, 2004], the Carrington storm in 1859 was approximately three times more intense than the 1989 storm [*Lakhina et al.*, 2005].

The main aims for the forecasting scheme is estimating the minimum value that the *Dst* index will reach and when it will happen. However, the knowledge of the remaining time for the magnetosphere to return to quiet time, or at least to 'non-dangerous time' is also an important output in which many technological systems rely on. These predictions are even more relevant for extreme geomagnetic storms.

Over the past years, it was assumed a proportional relationship between the decay rate of the ring current energy, and therefore of the *Dst* index, and the energy content of the ring current (through the Dessler-Parker-Sckopke relationship). Therefore, as a result of this linear dependence of the d*Dst*/d*t* on *Dst*, an exponential function was accepted for the recovery phase. Several authors [e.g., *Burton et al.*, 1975; *Hamilton et al.*, 1988; *Ebihara et al.*, 1998; *O'Brien & McPherron*, 2000; *Dasso et al.*, 2002; *Kozyra et al.*, 2002; *Wang et al.*, 2003; *Weygand & McPherron*, 2006; *Monreal MacMahon & Llop*, 2008] have dedicated much effort to find the decay time. In some cases, this was assumed to be a constant value [e.g. *Burton et al.* 1975] or dependent on the convective electric field $E_y$ [*O'Brien & McPherron*, 2000] or also dependent on the dynamic pressure [*Wang et al.*, 2003]. A highlight issue was the two-phase pattern (an early fast recovery followed by a slower one) observed in the *Dst* decay following intense geomagnetic storms, that was impossible to model assuming a unique exponential function [*Chapman*, 1952; *Akasofu et al.*, 1963; *Hamilton*, 1988; *Gonzalez et al.*, 1989; *Prigancova and Fel´Dshtein*, 1992; *Liemohn et al.*, 1999; *MacMahon and Llop,* 2008].

According to *Aguado et al.* [2010], the recovery phase of intense storms follows a hyperbolic decay, explaining in this way the entire recovery phase with one unique function dependent on two parameters: the minimum *Dst* value (*Dst*$_0$), which indicates the intensity of the storm and the moment when the recovery phase starts, i.e., *Dst* (*t* = 0) = *Dst*$_0$, and the recovery time ($\tau_h$), i.e., the time to get the value of *Dst*$_0$/2.

$$Dst(t) = \frac{Dst_0}{1 + \dfrac{t}{\tau_h}} \quad (1)$$

This semi-empirical model, which arises from a superposed epoch analysis of the recovery phase of intense geomagnetic storms in the period 1963-2003 with no significant injection of energy during this phase, states that the temporal variation of the *Dst* index is not proportional to *Dst*, but to *Dst*$^2$. Subsequently, the key issue of a hyperbolic decay model is that the recovery phase of the magnetosphere, as seen by *Dst* index, exhibits a nonlinear behavior.

Moreover, the two parameters included in the hyperbolic function, the minimum value of the *Dst* index and the recovery time, seem to be linearly related. This relationship was deduced after an analysis including only intense storms down to -400 nT. This paper presents a study of the recovery phase of the largest storms ever recorded [*Tsurutani et al.*, 2003] with two different aims: to check whether the hyperbolic model is able to reproduce properly observational data measured during the recovery phase of such extreme storms, and to investigate whether the recovery time and the intensity of the storm are still proportional for extreme storms.

This study is divided into five sections. Section 2 describes the data sets and processing. Section 3 contains methodology and results of the fitting of the recovery phases of all severe storms. Section 4 discusses the correlation between the intensity of the storm and the characteristic recovery time, both of them obtained as parameters from the fitting. Section 5 is a summary and discussion of the overall results obtained.

## 2. Data sets and data processing

The starting point for our study is the Table 1 of *Tsurutani et al.* [2003], which lists the "large magnetic storms" since 1857. We would like to check whether all twelve events listed in that table comply with the hyperbolic function (equation (1)). However, the *Dst* index is not available for all the events, as the International Geophysical Year 1957 was the starting date for the continuously computing of the *Dst* index at the World Data Center at Kyoto, Japan. As a result, only three events of the list of *Tsurutani et al.* [2003] have available *Dst* data. Before that date, only data from a number of observatories are available. For that reason, our first attempt was to estimate the *Dst* index from what we have called the 'Local Disturbance index' (*LDi*).

### 2.1 Data sets

The *Dst* index values and the horizontal (*H*) component of geomagnetic field with hourly resolution measured at each observatory used in this work are publicly available at the World Data Center (WDC) for Geomagnetism, Kyoto (http://wdc.kugi.kyoto-u.ac.jp/index.html). We have not found any available data at WDC for the events happened in 19th century or for the storm in October 1903 from Bombay observatory, but this storm was also a 'remarkable storm' at Potsdam observatory, and these data are available at WDC. Nevertheless, magnetic disturbances computed for mid-latitude stations might have a significant ionospheric component associated to the recorded activity which will make impossible to estimate the *Dst* index from magnetic field data from those stations. Therefore, we have removed from our study those events in Table 1 of *Tsurutani et al.* [2003] recorded at Potsdam (or the replacement stations, Seddin from 1908 through 1931, and Niemegk since 1932). For the event in September 1859, there are no data at WDC, but we have digitized data from Figure 3 of *Tsurutani et al.* [2003] from the Colaba (Bombay) magnetogram, which displays data for a two-day interval (1 Sep 16 h - 3 Sep 16 h Bombay local time). Table 1 displays the final list of events analyzed in this paper, including the observatory where data used were measured: seven severe storms out of the twelve events from the Table 1 of *Tsurutani et al.* [2003] are included.

**2.2 Data processing**

Most of the events of Table 1 were recorded by just one magnetometer. Therefore, to elaborate a global index, as the *Dst* index, from magnetometers distributed in longitude is not possible. Data processing made in this paper consists in obtaining a 'Local Disturbance index', i.e., an index (*i*) with local (*L*) information of the disturbance (*D*) during the storm time, from the *H* component of geomagnetic field measured at a determined observatory. The *LDi* index is obtained in a similar procedure to *Dst* [*Sugiura and Kamei*, 1991; *Häkkinen et al.*, 2003], but only from one geomagnetic observatory.

The first step is to define a baseline, $H_{baseline}$, for each storm and observatory. Our baseline consists in removing the periodic one-day variation and quiet time *H* value. Classification of days as 'quiet' or 'disturbed' is not available before 1932. Therefore, as we cannot consider the International Quietest Days (IQD) of the month for all the events in Table 1, we have set a procedure for obtaining the quietest days and remove the periodic variation as follows: we select the current month of the storm to determine the quietest days. We calculate the absolute value of the running difference for the hourly *H* data $|H(i+1)-H(i)|$. Next, we proceed to smooth $|H(i+1)-H(i)|$ with a 24-h window to find the minima. We should be aware that the window width does not alter the position of the minima, it just eliminates noise to visualize better the variation. The obtained minima will be our so-called quietest days. They are always selected avoiding discontinuities and recovery phases. Five quiet days, consecutive or not, are desirable in the selection. However, in some cases, only three days along the month can be considered as quiet days. The selection made in this way for the quietest days selected after 1932 coincides with the IQD except one of the days selected in April 1938. This discrepancy might arise because our selection is based on the *H* component of the geomagnetic field and the selection of the IQD is deduced from the *Kp* indices.

Once the quiet days are selected, they are averaged to form a 'quiet day model'. This one is replicated to create a synthetic periodic variation, i.e. the $H_{baseline}$. Then, the $H_{baseline}$ is subtracted from the original magnetogram signal. Top panel in Figure 1 displays the *H* component from Alibag observatory recorded in May 1921. The rectangle encloses the five quietest days selected for this event at this observatory. The *H* component after the $H_{baseline}$ removal (and therefore the daily variation and quiet time) is shown in bottom panel of Figure 1.

The hourly *LDi* index is finally obtained as $LDi(t) = \dfrac{H(t) - H_{baseline}}{\cos \varphi}$, where the cosine of the latitude of the magnetic observatory ($\varphi$) is used to normalize the index to the dipole equator. As a result, although the *LDi* misses the planetary perspective of the *Ds*t index, which average measurements widely spread in longitude, it can be still considered as a proxy of the disturbance at that specific station.

Previous researchers (e.g. *Akasofu and Chapman* [1964], *Chapman and Bartels* [1962], *Häkkinen et al.* [2003], *Moos* [1910] and *Bartels* [1932]) have revealed that magnetic disturbance at each observatory exhibits a diurnal variation, with greatest (least) storm-time disturbance at dusk (dawn). By examining the local time dependence of the disturbance time series for all observatories involved in the computation of *Ds*t index, *Love and Gannon* [2009] proposed a *Dst*-scalable local-time disturbance map, where what they called the local latitude-weighted disturbance (*Dlat*) was related to *Dst* index by a proportional relationship, i.e., $Dlat = \delta \cdot Dst$, being $\delta$ the following smooth function of local time, $\theta_h$, measured in continuous decimal hours:

$$\delta(\theta_h) = 0.9995 - 0.0149\cos(2\pi\frac{\theta_h}{24}) - 0.1803\sin(2\pi\frac{\theta_h}{24}) + \\ + 0.0157\cos(4\pi\frac{\theta_h}{24}) - 0.0130\sin(4\pi\frac{\theta_h}{24}) \quad (2)$$

Considering equation (2) and taking the *LDi* index as the local latitude-weighted disturbance, the *Dst* index is computed for those events of Table 1.

Figure 2 shows a comparison between the computed *Dst* index (solid black line) and the *Dst* index provided by the WDC (dashed gray line) for September 1957: (a) both indices as a function of time and (b) computed *Dst* versus WDC *Dst* and the linear regression (solid line). A similar comparison was done for February 1958 and March 1989, that is, for the months of the events where the *Dst* index is available. The good agreement between both indices ($r^2 > 0.93$) in all three cases reinforces the capability of computed *Dst*, when *Dst* is not available at WDC.

For the event on 2 September 1859, the coverage of available data is just two days. Therefore, the procedure described above for the establishment of the $H_{baseline}$ cannot be performed. During the 14 initial hours of data coverage, the *H* component value is almost constant and might be considered as representative of 'quiet time'. As a result, instead of selecting 'quiet days', we have computed the $H_{baseline}$ as the mean value of the *H* component during the interval 1 September 1859 16:00 to 2 September 1989 06:00 Bombay local time.

A similar approach was performed for the event of March 1989, as the day-night variation was already removed in those data provided through the World Data Center for Geomagnetism, Kyoto. For this event, the $H_{baseline}$ was computed as the mean value of the $H$ component for the IQD for March 1989. The computed $Dst$ index of every event is shown in Figure 3.

## 3. Fitting a hyperbolic function

A procedure to fit the recovery phase of the large magnetic storms is developed. We define the starting time ($t_0 = 0$) of the recovery phase of every storm at the time when the value of the computed $Dst$ index becomes minimum, and that corresponds to the peak of the storm at that observatory. Then we consider as recovery phase the first 48 hours after $t_0$, as after that time, it can be assumed that the magnetosphere is fully recovered. Considering a hyperbolic decay according to equation (1), the recovery time, $\tau_h$, for each of the seven recovery phases is computed using a standard least-squares procedure where the minimum value of the $Dst$ index ($Dst_{0,i}$) is also obtained as a parameter. Table 2 displays the results obtained for the two parameters (columns 5 an 6) and the correlation coefficient, $r^2$, (column 7) for every event in Table 1. The starting time for every event ($t_0$), chosen as $t_0 = 0$ for the fitting procedure, is shown in column 2 and the minimum value reached by the computed $Dst$ index appears in column 3. Column (4) includes the time interval after the starting time, $t_0$, included in the fitting procedure ($\Delta t$). In all cases a fitting including the first 48 hours of the recovery phase is performed. In two cases, which will be described below, a shorter interval is also fitted.

Figure 4 shows, as an example, one of the extreme geomagnetic events studied: the large storm that occurred in 1958 (event #6). This storm reached a peak value (computed $Dst_{min}$ = -475 nT) on 11 February 1958 at 12 UT. Circles in Figure 4 correspond to the computed $Dst$ values from the H component of the recorded magnetic field, while the solid line shows the fitted hyperbolic decay. The estimated hyperbolic decay time for this particular event is 8.4±0.9 hours, i.e., the magnetosphere has lost half of the injected energy from the solar wind 8.4 hours after its maximum disturbance.

From a visual inspection of the set of 7 storms we find that five of them (events #3 to #7) are in very good agreement with a hyperbolic decay, similar to the behavior shown in Figure 4. For these events, the correlation coefficient $r^2$ is larger than 0.87.

For the event #2 the correlation coefficient is low ($r^2 = 0.51$) and a three parameters hyperbolic decay function was needed to fit event #1In both cases we observe a systematic difference with respect to the fitted curve. In particular, additional peaks during recovery phases were visually and significantly observed (similar to the peaks reported by *Kamide et al.* [1998]). We guess that these storms probably received a significant energy input during the first 48 hours of their recovery and therefore, equation (1) is not suitable for the entire interval, as the injection of energy is not considered in the hyperbolic model. An example of this behavior is the storm in May 1921, which was recorded at two different stations: Alibag and Potsdam. In both cases, the *LDi* index, and therefore the *H* component recorded at those stations, decreases again around 20 hours after the starting time of the recovery phase, indicating that some injection of energy took place at that time (top panels in Figure 5). Therefore, we have assumed for this event a hyperbolic decay according to equation (1), but only for the first 20 hours of the recovery phase. Now, the event is in good agreement with a hyperbolic decay, as indicated by $r^2 > 0.96$ (see Figure 5 bottom panel).

The same situation happens again on the storm in September 1859 (event #1), but in this case a new decrease of ~300 nT in the *H* component of the terrestrial magnetic field takes place just 1.47 hours after the peak value of the storm. Because of the resolution of the data available for this period, the number of data to fit is similar to the previous events. The results show that the most severe storm ever registered, the Carrington event, is consistent during the first 90 minutes of the recovery phase with the hyperbolic model, with an $r^2$ value of 0.94 (Figure 6). Nevertheless, it should be noticed that the high resolution of computed *Dst* makes it more comparable to the *SYM-H* geomagnetic index than to the *Dst* index. Averaging the computed *Dst* to hourly resolution, the peak value obtained for this storm is -685 nT, which is comparable to the -640 nT reached during the 1989 storm responsible for the Hydro-Quebec power blackout, and 62% larger than the *Dst* peak value of the largest storm of solar cycle 23 happened in November 2003. This comparison indicates that although the Carrington storm seems to be the most intense geomagnetic storm ever recorded, it is not as extreme as usually is stated.

## 4. Correlations between fitting parameters

When fitting the hyperbolic model to intense geomagnetic storms, *Aguado et al.* [2010] obtained a linear relationship between the hyperbolic recovery time and the intensity of the storm. The fitting results from the previous section allow us to test whether both parameters ($\tau_h$ and $Dst_0$) are still proportional. Figure 7 displays the parameter $\tau_h$ versus $Dst_0$ for all of the events analyzed in this paper. The parameters from the fitting to the four "mean recovery phases" by *Aguado et al.* [2010] are also displayed in the plot as gray triangles. Figure 7 evidences that the linear relationship is not suitable for extreme storms, although the parameters from *Aguado et al.* [2010] appear to follow the tendency of those in this study. The whole set of data fits to an exponential growth given by $\tau_h(\text{h}) = (21 \pm 3)\exp\left[(2.4 \pm 0.5) \times 10^{-3} Dst_0(\text{nT})\right]$, with a correlation coefficient $r^2 = 0.84$ (solid line in Figure 7). When this exponential function is expanded by the first two terms of the Taylor series expansion, this result is consistent with the expression proposed by *Aguado et al.* [2010].

## 5. Summary and Discussion

For the first time we reproduce by an empirical function the fast recovery of the magnetometer records during the most severe storms ever registered, as the Carrington event. After introducing a local index to quantify the disturbance registered at a determined observatory, the *LDi* index, which is useful to estimate *Dst* when there is no global index available, this study shows the high accuracy of the hyperbolic decay function to reproduce the recovery phase of the magnetosphere after an extreme storm. Our results show that this function can easily achieve the fast recovery that takes place after the extremely negative *Dst* values extending the range where the model proposed by *Aguado et al.* [2010] is applicable.

As stated by *Aguado et al.* [2010], the hyperbolic decay function is able to provide by a unique continuous function a steep rise in the early recovery phase and a smooth one in the late phase. Therefore, the high accuracy of the hyperbolic fitting in reproducing the recovery phase of *Dst* index in extreme storms addresses the existence of diverse processes of different nature (flow-out, charge exchange of different ions, particle precipitation by wave-particle interaction, etc.) involved in a gradual way. The outcome is a non-constant degree of reduction of *Dst* index (defined as $-(dDst/dt)/Dst$) and, as a consequence, a non-linear coupling of *dDst/dt* upon *Dst*. These results are a key point in magnetospheric physics, as a hyperbolic decay function for the recovery phase means that the losses of energy in the magnetosphere are proportional to the square of the energy content, instead of proportional to the energy content itself, as indicated by an exponential decay.

As an additional point to the goodness of the hyperbolic function to reproduce the recovery of the magnetic field measured at terrestrial surface after a severe disturbance, the results of this study demonstrate that the time that takes the magnetosphere to recover depends in an exponential way on the intensity of the storm, as indicated by the relationship between the two parameters involved in the hyperbolic decay function, i.e., the hyperbolic recovery time and the minimum value of the index used to quantify the disturbance. The exponential function obtained is consistent with the linear function proposed by *Aguado et al*. [2010] when the severity of the storm diminishes.

Despite the goals of this study, we should notice that fully understanding severe magnetic field disturbances measured at terrestrial surface during the early recovery phase of severe storms relies critically on solar wind transients. Two out of the seven recovery phases of severe storms analyzed present double peaks in the *Dst* index, which are assumed as additional injection of energy, although the lack of interplanetary magnetic field data for those dates avoids the certainty on that assumption.

The hourly resolution of data is also a limitation in the analysis of some events. High resolution data, when available as for the Carrington event, illustrates that during the first hour the recorded disturbance can be properly reproduced by the hyperbolic function, even when a second peak appears in the very early recovery phase (less than two hours after the maximum disturbance).

**Acknowledgements**

This work has been supported by grants PN-AYA2009-08662 from the Comisión Interministerial de Ciencia y Tecnología (CICYT) of Spain and PPII10-0183-7802 from the Junta de Comunidades de Castilla-La Mancha of Spain. We also thank the data from the World Data Center for Geomagnetism, Kyoto, and to the geomagnetic observatories which provide the data through that data center.

**References**

Aguado, J., C. Cid, E. Saiz, and Y. Cerrato (2010), Hyperbolic decay of the Dst Index during the recovery phase of intense geomagnetic storms, *J. Geophys. Res.*, *115*, A07220, doi:10.1029/2009JA014658.

Akasofu, S. I. and Chapman, S. (1964), On the asymmetric development of magnetic storm fields in low and middle latitudes, *Planet. Space Sci.*, *12*, 607–626.

Akasofu, S. I., S. Chapman, and B. Venkatesan (1963), The Main Phase of Great Magnetic Storms, *J. Geophys. Res., 68*, 3345-3350.

Bartels, J. (1932), Terrestrial-magnetic activity and its relations to solar phenomena, *Terr. Magn. Atmos. Electr.*, *37*, 1–52.

Burton, R.K., R.L. McPherron, and C.T. Russell (1975), An empirical relationship between interplanetary conditions and Dst, *J. Geophys. Res.*, *80*, 4204-4214.

Chapman, S. (1952), The Earth's Magnetism, *American Journal of Physics 20*, 316-316.

Chapman, S. and Bartels, J. (1962), Geomagnetism, Volumes 1 & 2, Oxford Univ. Press, London, UK.

Cliver, E. W., and L. Svalgaard (2004), The 1859 Solar–Terrestrial disturbance and the current limits of extreme Space Weather Activity, *Sol. Phys.*, *224*, nº 1, 407-422, doi: 10.1007/s11207-005-4980-z.

Dasso, S., D. Gómez, and C. H. Mandrini (2002), Ring current decay rates of magnetic storms: A statistical study from 1957 to 1998, *J. Geophys. Res.*, *107(A5)*, 1059, doi:10.1029/2000JA000430.

Ebihara, Y., and M. Ejiri (1998), Modeling of solar wind control of the ring current buildup: A case study of the magnetic storms in April 1997, *Geophys. Res. Lett.*, *25(20)*, 3751–3754, doi:10.1029/1998GL900006.

Echer, E., Gonzalez,W.D., Tsurutani,B.T. (2008), Interplanetary conditions leading to superintense geomagnetic storms (Dst ≤ -250 nT) during solar cycle 23, *Geophys. Res.Lett.*, *35*, L06S03.doi:10.1029/2007GL031755.

Gonzalez, W. D., A. L. C. Gonzalez, B. T. Tsurutani, E. J. Smith, and F. Tang (1989), Solar wind-magnetosphere coupling during intense magnetic storms (1978-1979), *J. Geophys. Res., 94*, 8835-8851.

Gonzalez, W. D., J. A. Joselyn, Y. Kamide, H. W. Kroehl, G. Rostoker, B. T. Tsurutani, and V. M. Vasyliunas (1994), What is a geomagnetic storm?, *J. Geophys. Res.*, *99*, 5571–5792, doi:10.1029/93JA02867.

Gonzalez, W.D., Tsurutani, B.T., Lepping, R.P., and Schwenn, R. (2002), Interplanetary

phenomena associated with very intense geomagnetic storms. *J. Atmos. Sol. Terr. Phys., 64*, 173–181.

Häkkinen, L. V. T., I. Pulkkinen, R. J. Pirjola, H. Nevanlinna, E. I. Tanskanen, N. E. Turner (2003), Seasonal and diurnal variation of geomagnetic activity, revised Dst vs. external drivers, *J. Geophys. Res.*, *108(A2)*, 1060, doi:10.1029/2002JA009428

Hamilton, D. C., G. Gloeckler, F. M. Ipavich, W. Stüdemann, B. Wilken, and G. Kremser (1988), Ring current development during the great geomagnetic storm of February 1986, *J. Geophys. Res.*, *93(A12)*, 14,343–14,355, doi:10.1029/JA093iA12p14343.

Kozyra, J. U., M. W. Liemohn, C. R. Clauer, A. J. Ridley, M. F. Thomsen, J. E. Borovsky, J. L. Roeder, V. K. Jordanova, and W. D. Gonzalez (2002), Multistep Dst development and ring current composition changes during the 4–6 June 1991 magnetic storm, *J. Geophys. Res.*, *107(A8),*1224, doi:10.1029/2001JA000023.

Lakhina, G. S., S. Alex, B. T. Tsurutani, and W.D. Gonzalez (2005), ―Research on Historical Records of Geomagnetic Storms, in *Proceedings IAU Symposium*, *226,* K. P. Dere, J. Wand, and Y. Yan, eds., doi:10.1017/S1743921305000074.

Liemohn, M. W., J. U. Kozyra, V. K. Jordanova, G. V. Khazanov, M. F. Thomsen, and T. E. Cayton (1999), Analysis of early phase ring current recovery mechanisms during geomagnetic storms, *Geophys. Res. Let..,* *26*, 2845-2848.

Love, J. J., and J. L. Gannon (2009), Revised *Dst* and the epicycles of magnetic disturbance: 1958–2007, *Ann. Geophys.*, *27*, 3101-3131.

Monreal MacMahon, R., and C. Llop (2008), Ring current decay time model during geomagnetic storms: A simple analytical approach, *Ann. Geophys.*, *26*, 2543–2550.

Moos, N. A. F. (1910), Colaba Magnetic Data, 1846 to 1905. Part I: Magnetic Data and Instruments; Part II: The Phenomenon and its Discussion, Government Central Press, Bombay, India.

O'Brien, T. P., and R. L. McPherron (2000), An empirical phase space analysis of ring current dynamics: Solar wind control of injection and decay, *J. Geophys. Res., 105(A4)*, 7707–7719, doi:10.1029/1998JA000437.

Prigancova, A., and I. A. Fel'Dshtein (1992), Magnetospheric storm dynamics in terms of energy output rate, *Planetary and Space Science, 40,* 581-588.

Sugiura, M., and T. Kamei (1991), Equatorial Dst index, 1957- 1986, *Int. Assoc. Geomagn. Aeron. Bull.*, *40*, 246.

Tsurutani, B. T., W. D. Gonzalez, F. Tang, and Y. T. Lee (1992), Great magnetic storms, *Geophys. Res. Lett.*, *19*, 73.

Tsurutani, B. T., W. D. Gonzalez (1987), "The cause of high-intensity long-duration continuous AE activity (HILDCAAs): Interplanetary Alfvén wave trains," *Planetary and Space Science*, 35, 405-412.

Tsurutani, B. T., W. D. Gonzalez, G. S. Lakhina, and S. Alex (2003), The extreme magnetic storm of 1–2 September 1859, *J. Geophys. Res.*, *108(A7)*, 1268, doi:10.1029/2002JA009504.

Wang, C. B., J. K. Chao, and C.-H. Lin (2003), Influence of the solar wind dynamic pressure on the decay and injection of the ring current, *J. Geophys. Res.*, *108(A9)*, 1341, doi:10.1029/2003JA009851.

Weygand, J. M., and R. L. McPherron (2006), Dependence of ring current asymmetry on storm phase, *J. Geophys. Res.*, *111*, A11221, doi:10.1029/2006JA011808.

**Table 1.** Chronological list of large geomagnetic storms analyzed in this paper

| Event # | Year | Month | Day | Observatory | H range (nT) | Geomagnetic latitude[a] |
|---|---|---|---|---|---|---|
| 1 | 1859 | September | 1-2 | Bombay | 1720 | 9.74 |
| 2 | 1921 | May | 13-16 | Alibag | >700 | 9.46 |
| 3 | 1928 | July | 7 | Alibag | 780 | 9.45 |
| 4 | 1938 | April | 16 | Alibag | 530 | 9.37 |
| 5 | 1957 | September | 13 | Alibag | 580 | 9.29 |
| 6 | 1958 | February | 11 | Alibag | 660 | 9.29 |
| 7 | 1989 | March | 13 | Kakioka | 640 | 26.6 |

[a] Geomagnetic latitude for all observatories have been computed for the closest year to the event that was available using the transformation offered by the WDC for Geomagnetism, Kyoto at http://wdc.kugi.kyoto-u.ac.jp/igrf/gggm/index.html.

**Table 2.** Parameters obtained after fitting a hyperbolic function to the events of Table 1. See text for details.

| Event # | $t_0$ (yyyy mm dd hh:mm) | $Dst_c$ min (nT) | $\Delta t$ (h) | $\tau_h$ (h) | $Dst_0$ (nT) | $r^2$ |
|---|---|---|---|---|---|---|
| 1 | 1859 09 02 10:15[b] | -1697 | 48 | 0.10±0.02 | -1600±135 | 0.68 |
|   |   |   | 1.47 | 0.14±0.02 | -1753±103 | 0.93 |
| 2 | 1921 05 15 05 | -713 [c] | 48 | 7.27±1.70 | -646±73 | 0.51 |
|   |   |   | 20 | 3.55±0.34 | -767±32 | 0.96 |
| 3 | 1928 07 08 10 | -506 | 48 | 4.55±0.45 | -585±32 | 0.88 |
| 4 | 1938 04 16 10 | -263 | 48 | 6.46±0.65 | -267±13 | 0.91 |
| 5 | 1957 09 13 10 | -532 | 48 | 3.67±0.30 | -541±22 | 0.93 |
| 6 | 1958 02 11 11 | -475 | 48 | 8.40±0.90 | -457±23 | 0.87 |
| 7 | 1989 03 14 00 | -674 | 48 | 6.11±0.61 | -688±34 | 0.88 |

[b]This event was fitted to a three parameters hyperbolic decay function, i.e. $LDi(t) = \dfrac{LDi_0}{1+\dfrac{t}{\tau_h}} + C$, where the result for the third parameter is $C = -165 \pm 15$ nT.

[c]There is a data gap the hour before $t_0$.

**FIGURE 1**

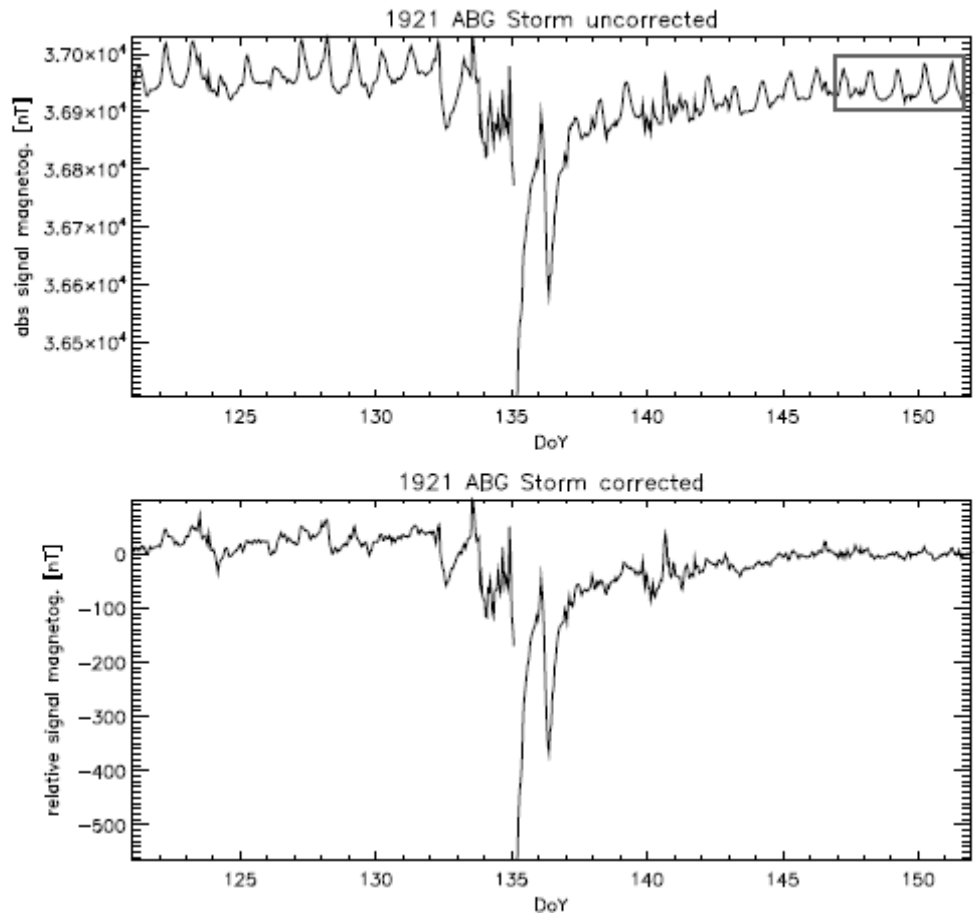

**FIGURE 2**

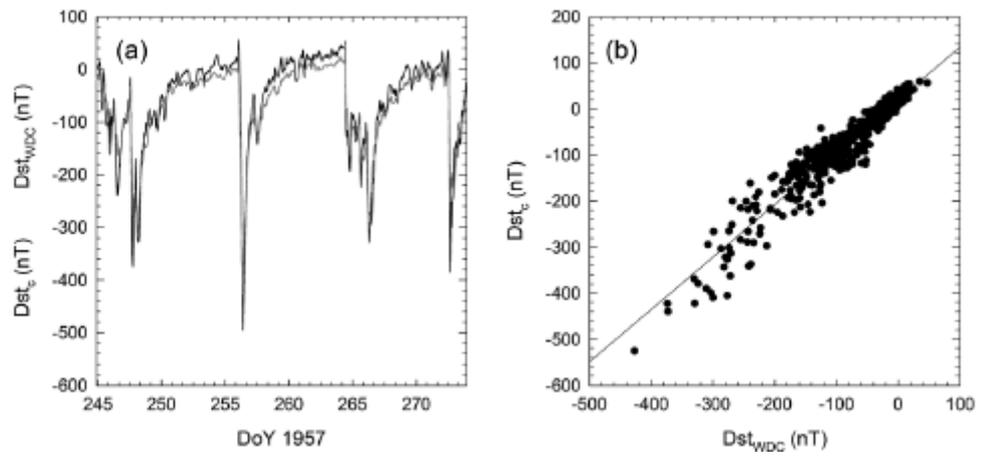

**FIGURE 3**

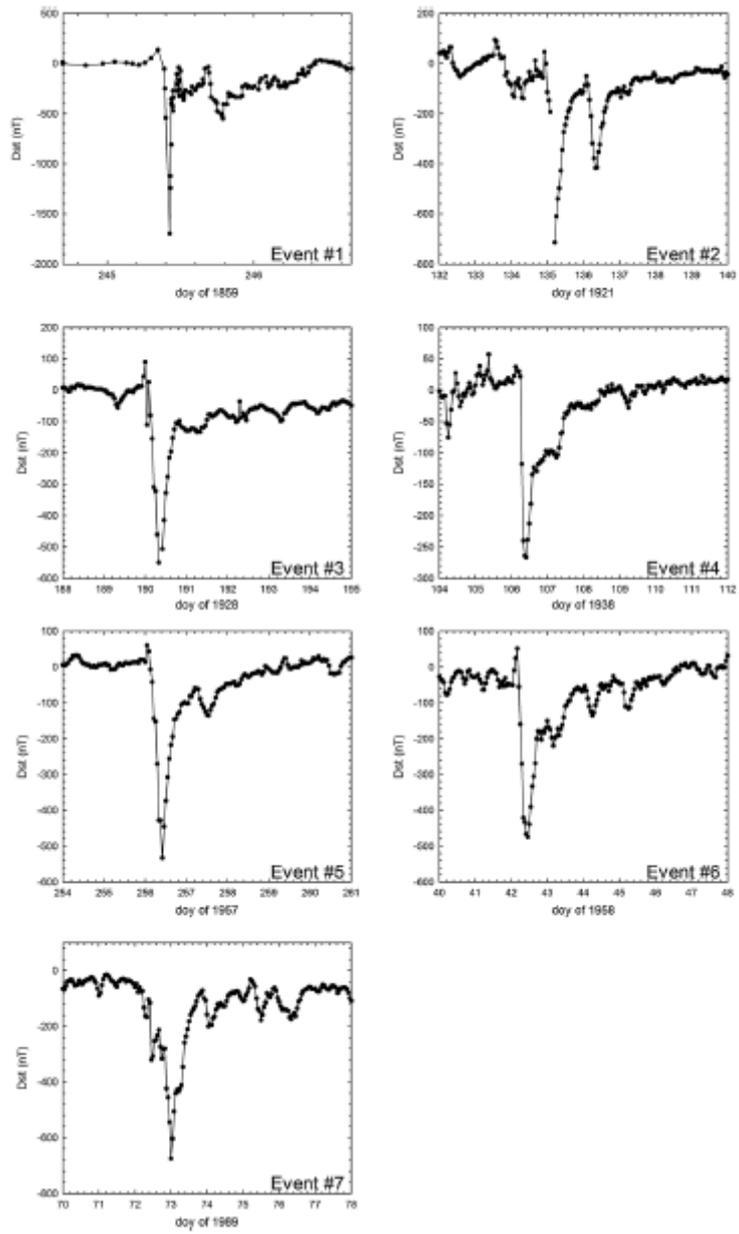

**FIGURE 4**

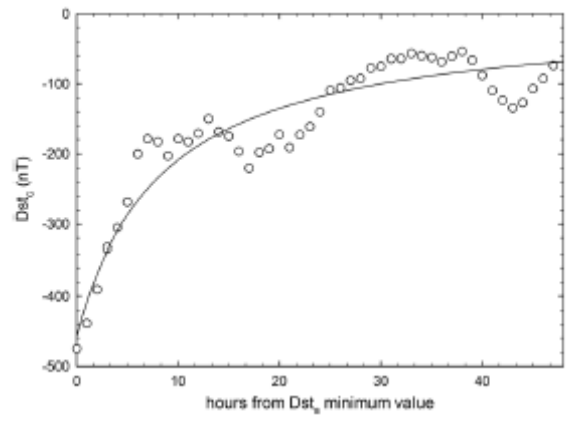

**FIGURE 5**

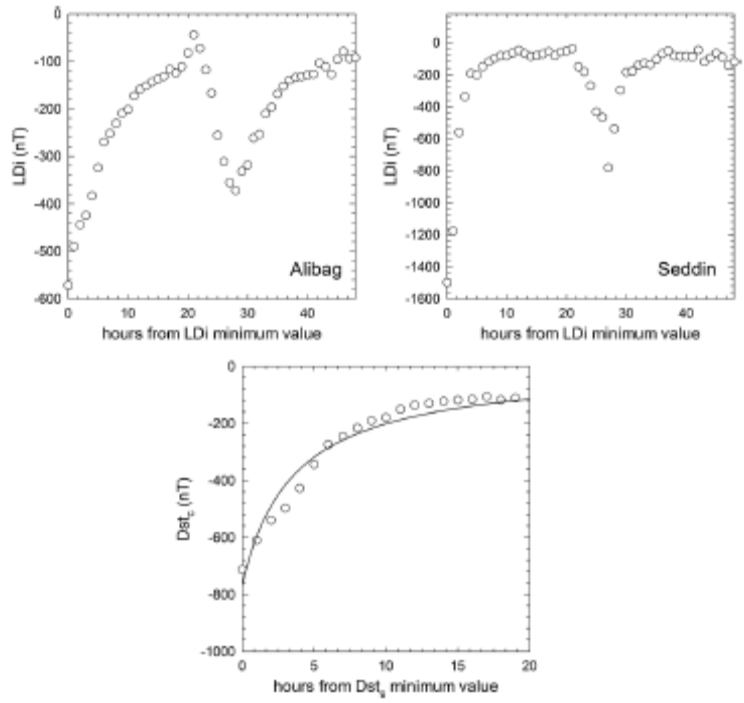

**FIGURE 6**

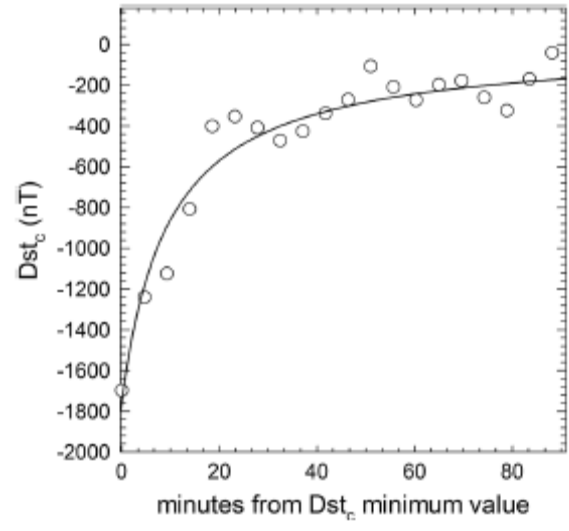

**FIGURE 7**

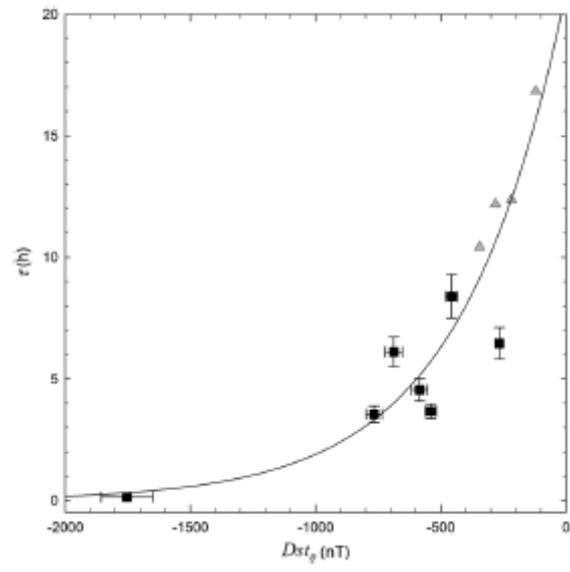